# TESTING RESULTS OF THE PROTOTYPE BEAM ABSORBER FOR THE PXIE MEBT

C. Baffes[#], A. Shemyakin, Fermilab[*], Batavia, IL 60510, USA

*Abstract*

One of the goals of the PXIE program at Fermilab [1] is to demonstrate the capability to form an arbitrary bunch pattern from an initially CW 162.5 MHz H- bunch train coming out of an RFQ. The bunch-by-bunch selection will take place in the 2.1 MeV Medium Energy Beam Transport (MEBT) [2] by directing the undesired bunches onto an absorber that needs to withstand a beam power of up to 21 kW, focused onto a spot with a ~2 mm rms radius. A prototype of the absorber was manufactured from molybdenum alloy TZM, and tested with an electron beam up to the peak surface power density required for PXIE, 17W/mm$^2$. Temperatures and flow parameters were measured and compared to analysis. This paper describes the absorber prototype and key testing results.

## PXIE ABSORBER CONCEPT

The overall length of the PXIE absorber is constrained to 650mm by the optics design of the MEBT. In order to limit the power density on the absorber, the absorbing surface is inclined such that incident beam strikes at a grazing angle of incidence (29mrad). This geometry results in approximately 25% of the incoming H- beam power being reflected. The peak absorbed surface power density is 17 W/mm$^2$.

In order to combat the driving gas load of recombined H$_2$ and maintain vacuum better than 10$^{-6}$ Torr within the absorber enclosure, turbo pumps with combined pumping speed of 2000 l/s are planned. The absorbing surface is implemented in Molybdenum alloy TZM, which combines good high-temperature mechanical properties and a resistance to beam-induced blistering. The absorbing surface is longitudinally segmented to relieve thermally-induced stresses. Additional details of the PXIE concept may be found in [3] and [4].

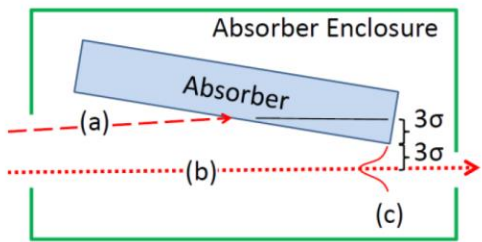

Figure 1: Schematic representation of the MEBT absorber, showing (a) chopped beam, (b) passed beam, (c) 6σ shift between the chopped and passed beams

## PROTOTYPE DESIGN

A prototype program was developed to retire the perceived risks of the PXIE absorber concept, specifically the complicated fabrication process, high surface power density, associated aggressive thermal conditions, and accuracy of analysis. In this design concept, longitudinal heat transfer is interrupted by stress relief slits (see Figure 2), and so heat flows primarily transversely. As a result, it is possible to replicate PXIE-like thermal conditions in a relatively short longitudinal space: the prototype's absorbing surface is 116mm long. Coolant flows transversely through the absorber body in 300μm wide cooling channels fabricated by EDM. TZM components were joined by brazing, and interfaced to stainless plumbing through a series of material transitions. Temperatures within the absorber body are monitored by an array of thermocouples. A more comprehensive description of the prototype, testing methods and results is given in [5].

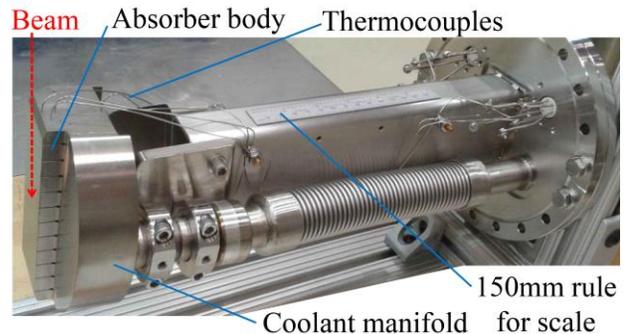

Figure 2: Absorber Prototype

## TEST RESULTS

*Equipment, Instrumentation and Methods*

A dedicated test stand was prepared for this program. The prototype absorber can be heated by an 27.5keV 0.2A (max) electron beam. A significant portion of the beam energy (~55%) is reflected or carried away in secondary electrons, so the maximum power than can be absorbed in the prototype is 2.5kW. The beam size can be adjusted by solenoids, position can be adjusted by corrector dipoles.

The beam interaction with the surface produces bright visible Optical Transition Radiation light (OTR), and beam heating of the surface can produce visible thermal radiation. Light of either type passes through a quartz vacuum viewport and is imaged by a digital camera. The intensity of OTR may be used to understand the beam profile, which contained non-uniformities which required inclusion in thermal analysis. The intensity of thermal radiation was used to reconstruct surface temperature profiles. A variety of optical filters of varying central wavelength were used to separate OTR (broad band and light intensity linear with current) from thermal radiation (typically longer wavelength and light intensity strongly non-linear with surface temperature). These methods are further described in [5].

In addition to thermometry in the absorber body, coolant temperature and flow rates were recorded. This permitted for rough calorimetry to understand the energy deposition to the absorber.

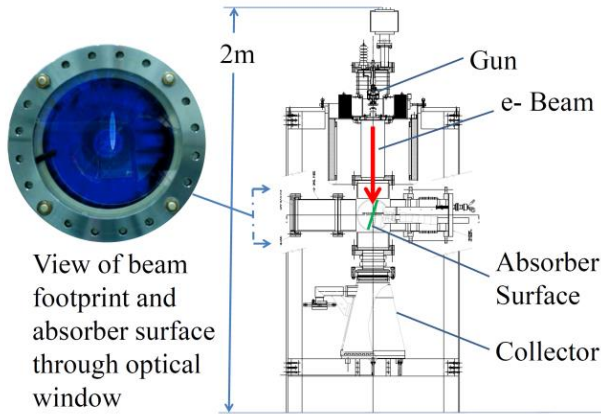

Figure 3: The test stand and OTR beam image

*Energy Deposition and Thermal Performance*

Over several months of testing, beam current was increased and beam size was decreased until the peak power density expected for PXIE of 17W/mm$^2$ was achieved over one "fin", a fin being an area of the surface bordered and thermally isolated by stress relief slits. In this condition, peak surface temperatures of ~1300K were reconstructed from optical measurement and analysis. After exposure to this thermal condition, the absorber survived and did not exhibit any symptoms of damage. This is the primary result of this testing program.

Local heating effects of two types were observed. The absorber surface exhibited a few permanent "hot spots," areas with characteristic size <1mm$^2$ that were heated even with low power density. It is speculated that these hot spots were dust or similar contamination in poor thermal contact with the absorber surface. A second, more significant type of local heating was the result of fine structure within the beam, as revealed by OTR. For purposes of analysis correlation, data were taken at two different focusing conditions, a tight focusing condition producing average power density of 17W/mm$^2$, and an intermediate focusing condition producing average power density of 10W/mm$^2$. In the intermediate focusing condition, clean separation of OTR and thermal radiation was possible. However, at the tight focusing condition, attempts to understand structure within the beam were hampered by the fact that thermal radiation was present even at the short-wavelength limit (~450nm) of the camera/filter system. As such, it was necessary to estimate the beam profile in the tight focusing condition by scaling the profile in the intermediate focusing condition, shown in Figure 4.

For each focusing condition, visible radiation and thermocouple temperature profiles were captured for comparison with analysis.

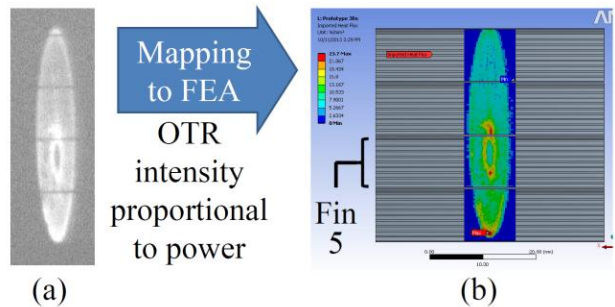

Figure 4: Reconstruction of energy deposition at intermediate focusing condition (a) – Blue filtered OTR image (b) – Analyzed beam profile. Power density on Fin 5 10 W/mm$^2$, 25 W/mm$^2$ peak

*Thermal Analysis Correlation*

Given the OTR-derived beam profile and total energy deposition estimated from calorimetry, finite element analysis was performed in ANSYS [6] to predict TZM temperatures, both at the surface and at the discrete thermocouple locations within the absorber body.

Predicted surface temperatures were compared to temperature estimates calculated from thermal radiation. As described in [5], fitting parameters used in the radiation calculations were adjusted to optimize fit with FEA simulation, so any agreement better than the inherent uncertainty of the optical measurement (±150K) is, to some extent, enforced. However, some encouragement is taken from the similarity with which analysis and measurement reconstruct fine structure within the temperature profile. An example comparison is shown in Figure 5.

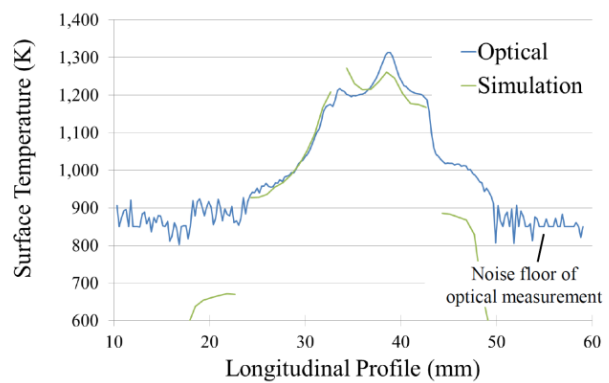

Figure 5: Comparison of FEA simulation and optical measurement of surface temperature along a linear profile at maximum power density

Thermocouple measurements were also compared to simulation. After correcting for systematic errors associated with thermocouple mounting (see [5]), simulations agreed well with measurements, and tended to under-predict the observed temperature rise by 3-10%.

*Cooling Studies*

In order to provide aggressive cooling of the absorber, 300μm wide cooling channels were machined directly in the TZM. The narrow width of the channel was intended to create laminar flow conditions, with relatively constant thermal performance over a range of flow rates. Simulation was performed to optimize channel design, specifically to ensure uniform flow distribution to all channels and avoid boiling conditions.

In order to test the cooling design, energy depostion was established in the tight focusing/maximim power density regime. Flow was incrementally increased such that transverse velocity varied between 0.6 and 3.5m/s. Thermocouple temperatures were monitored to detect changes in the cooling effectiveness. Over this range, no drastic inflections indicative of regime change (e.g. boiling) were observed. For flow velocities above 1m/s there was a very slight improvement in cooling (relative to the expectation of invariant laminar heat tranfer). This may indicate the onset of the transition to turbulence, which would be helpful in this application.

## CONCLUSIONS AND NEXT STEPS

A prototype of the PXIE MEBT beam absorber has been built, and tested using an electron beam to an average absorbed power density of 17W/mm$^2$. This is representative of PXIE operating at the full 10mA beam current. The prototype survived the testing, and exhibited thermal performance consistent with analysis. The choice of Molybdenum TZM as a high-temperature absorbing material was validated.

Even so, the fabrication and testing cycle revealed some deficiencies in the design. Better management of reflected energy is desirable. Having water flow through the inherently brittle TZM required a complex design and fabrication, and created the possibility of water-to-vacuum failure mode. Given the good thermal performance of the prototype, the authors were emboldened to sacrifice some thermal performance in favor of a simpler design. This design relies on thermal contact between the TZM absorbing surface and a cooling block made of aluminum. A cross section is shown in Figure 6.

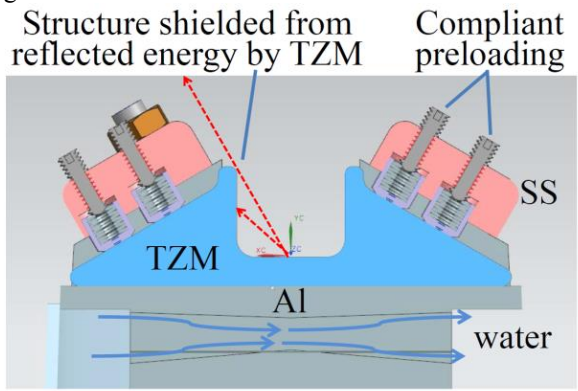

Figure 6: 2$^{nd}$ generation prototype cross section. Beam is into page, side walls contain and direct reflected energy

At the preloaded joint between TZM and aluminum, thermal contact is enhanced by a compliant graphite interface foil.

In this concept, the TZM absorbing surface extends to include side "walls." These walls reabsorb some of the particles reflected from the absorber surface, and in the H- PXIE application limit the areas of the absorber enclosure where cooled and bilstering-resistant secondary absorbing surface must be provided.

This 2$^{nd}$ generation prototype is currently being assembled at Fermilab. Thermal testing in the test bench electron beam is expected to commence in summer of 2014. If test results are favorable, the further tests of resistance to H- induced blistering will be conducted in the PXIE beam line.

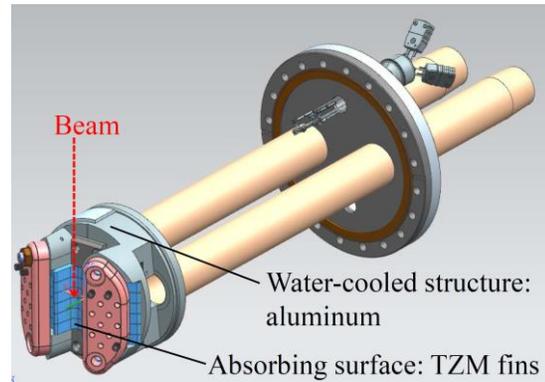

Figure 7: 2$^{nd}$ generation prototype, comprised of six TZM fins with longitudinal length of 1cm

## ACKNOWLEDGMENTS

The authors wish to acknowledge the efforts of K. Carlson, C. Exline, B. Hanna, A. Mitskovets, L. Prost, R. Thurman-Keup and J. Walton in assembling, commissioning and operating the test stand used to produce these data.